\documentstyle[12pt,aaspp4,psfig]{article} 
 
\begin{document}
 
\title{The 9.7 Micron Silicate Dust Absorption Toward the Cygnus A
Nucleus and the Inferred Location of the Obscuring Dust
\altaffilmark{1}} 
\author{Masatoshi Imanishi \altaffilmark{2}}  
\affil{National Astronomical Observatory, Mitaka, Tokyo 181-8588, Japan}
\author{and \\
Shiro Ueno}
\affil{Space Utilization Research Program, Tsukuba Space Center, 
National Space Development Agency of Japan,  
2-1-1 Sengen, 
Tsukuba 305-8505, Japan}
\altaffiltext{1}{
Data presented here were obtained at the W. M. Keck Observatory, 
which is operated as a scientific partnership among the 
California Institute of Technology, the University of California, and 
the National Aeronautics and Space Administration.                
The Observatory was made possible by the generous financial support 
of the W. M. Keck Foundation.}
\altaffiltext{2}{Present address: Institute for Astronomy, 
University of Hawaii, 2680 Woodlawn Drive, Honolulu, Hawaii 96822, 
USA}

\begin{abstract} 
 
We report the detection of a 9.7 $\mu$m silicate dust absorption
feature toward the Cygnus A nucleus.  Its optical depth is, however,
significantly smaller than that expected from the dust extinction
toward the background $L$-band emission region ($A_{\rm V}$ $\sim$
150 mag).  We argue that the most likely explanation for the small
optical depth is that the obscuring dust exists so close to the
central engine that a temperature gradient occurs. 
Our calculation confirms that the small optical
depth and the spectral energy distribution at 3--30 $\mu$m can be
quantitatively reproduced by this explanation.  Combining this picture
with the huge AGN luminosity, the emission properties of Cygnus A are
consistent with those of a type 2 quasar, that is, a highly luminous
AGN that is highly obscured ($A_{\rm V}$ $>$ 50 mag) by a dusty
torus with an inner radius of $<$10 pc, and not by $>$ a few 100 pc scale
dust in the host galaxy.

\end{abstract} 
 

\section{Introduction} 

A wide range of the observed properties of active galactic nuclei
(AGNs) can be explained by differing viewing angles toward a dusty
torus with an inner radius of $<$10 pc around an accreting 
supermassive blackhole (Antonucci 1993).  However, an important question
about AGNs that has not yet been answered is ``how common are type 2
 quasars?'' where we define type 2 quasars as highly luminous AGNs
(bolometric luminosity of $>$10$^{12}L_{\odot}$, or 2--10 keV hard
X-ray luminosity of $>$10$^{44}$ ergs s$^{-1}$)
%--------------
\footnote{
We adopt $H_{\rm 0}$ = 75 km s$^{-1}$ Mpc$^{-1}$ and $q_{\rm 0}$ = 0.5 
throughout this paper.} 
%-------------- 
that are highly obscured ($A_{\rm V}$ $>$ 50 mag) by a dusty torus.
To explain a significant fraction of the huge infrared
luminosity of ultra-luminous infrared galaxies ($L_{\rm IR}$ $>$
10$^{12}L_{\odot}$; Sanders \& Mirabel 1996) by dust emission 
powered by obscured AGN activity, 
the presence of such type 2 quasars is required.  However,
it has been suggested that dust around a highly luminous AGN is quickly
expelled
by strong radiation pressure and outflow activities,
and therefore virtually no type 2 quasars exist 
(Halpern, Turner, \& George 1999).

One way to search for a type 2 quasar is to find an X-ray source
with a large hard X-ray to soft X-ray flux ratio (i.e., an indication of
soft X-ray attenuation by absorption), and then perform optical follow-up
spectroscopy to derive the redshift, and hence estimate the
intrinsic hard X-ray luminosity (e.g., Sakano et al. 1998).  If both
the intrinsic 2--10 keV hard X-ray luminosity and X-ray absorption
are estimated to be high (e.g., $L_{\rm X}$(2--10 keV) $>$ 10$^{44}$
ergs s$^{-1}$ and $N_{\rm H}$ $>$ 10$^{23}$ cm$^{-2}$), the source is
a candidate type 2 quasar.  
However, X-ray absorption ($N_{\rm H}$) is caused both by gas and dust, 
and the $N_{\rm H}$/$A_{\rm V}$ 
ratio toward an AGN is often higher than that of the Galaxy 
($N_{\rm H}$/$A_{\rm V}$ = 1.8 $\times$ 10$^{21}$ cm$^{-2}$ mag$^{-1}$; 
Predehl \& Schmitt 1995), as much as 10 times higher in some cases 
(Alonso-Herrero, Ward, \& Kotilainen 1997; Simpson 1998).
Hence, the estimation of dust obscuration ($A_{\rm V}$) from hard X-ray
absorption is highly uncertain.  If the optical spectrum has
sufficient spectral resolution and sensitivity, we can investigate
dust obscuration by looking for the absence of a broad ($>$ 2000 km
s$^{-1}$ in full width at half-maximum; FWHM) component of the
rest-frame optical hydrogen emission lines (e.g., Ohta et al. 1996;
Georgantopoulos et al. 1999).  Its absence, however, indicates only 
that dust obscuration is at least several mag in $A_{\rm V}$.  Such a
small amount of dust is insufficient to explain the huge amount of 
infrared emission by dust heated by AGN activity.  
Therefore, the question
as to whether there are type 2 quasars that can explain the huge 
infrared luminosity by dust heated by obscured AGN activity cannot 
be answered explicitly by this method.

Study of thermal infrared regions (3--30 $\mu$m) is a powerful tool 
in the search for such a type 2 quasar.  
Firstly, since thermal infrared
emission is visible through high dust obscuration, as high as $A_{\rm
V}$ $>$ 50 mag, we can find a sample of highly luminous and highly
dust obscured ($A_{\rm V}$ $>$ 50 mag) AGNs.  Secondly, we can examine
the location of the obscuring dust.  Obscuration can be either by dust
in a dusty torus in the vicinity of a central engine ($<$10 pc in
inner radius), or by dust in the host galaxy ($>$ a few 100 pc scale).  
Needless to say, dust must be present in the location of the former 
in order for it to be heated by AGN activity.  
We can distinguish between these two possibilities by
looking for the presence of a temperature gradient in the obscuring
dust.  If a dusty torus is responsible for the
obscuration, a temperature gradient is predicted to occur,
with the temperature of the dust decreasing with increasing 
distance from the central engine (Pier \& Krolik 1992).  
The temperature of the
innermost dust is expected to be $\sim$1000 K, close to the dust
sublimation temperature.  Since emission at 3 $\mu$m is dominated by
dust at $\sim$1000 K, the extinction estimated using the $\sim$3
$\mu$m data should reflect the value toward the innermost dust around
the central engine.  On the other hand, the extinction estimated using
$\sim$10 $\mu$m data should be lower, because the dust at $\sim$300 K,
a dominant emission source at $\sim$10 $\mu$m, is located further out 
than the $\sim$1000 K dust, and thus the $\sim$10 $\mu$m data can only
trace the extinction toward the outer region.  In fact, if there is a
temperature gradient, the optical depth of the 9.7 $\mu$m silicate
dust absorption is predicted to be lower than the actual column
density toward the central engine (Pier \& Krolik 1992).  
Such a temperature gradient is not thought to occur in 
the $>$ a few 100 pc scale dust in
the host galaxy.  Hence, comparing extinction estimates at $\sim$3
$\mu$m and $\sim$10 $\mu$m can provide useful information on the
location of obscuring dust.

Cygnus A (3C 405; $z$ $=$ 0.056) has a highly luminous radio-loud 
AGN with a bolometric luminosity of $>$10$^{45}$ ergs s$^{-1}$ (Stockton,
Ridgway, \& Lilly 1994) and with an extinction-corrected 2--10 keV
hard X-ray luminosity of 1--5 $\times$ 10$^{44}$ ergs s$^{-1}$
(Ueno et al. 1994; Sambruna, Eracleous, \& Mushotzky 1999).  The dust
extinction toward the background $L$-band ($\sim$3.5 $\mu$m) emission
region of the nucleus is estimated to be $A_{\rm V}$ $\sim$ 150 mag, 
based on the comparison of the {\it observed} $L$-band luminosity 
to the predictions from the optical [OIII] emission line and 
the extinction-corrected 2--10 keV hard X-ray luminosities (Ward 1996).  
Both results suggest that Cygnus A is a candidate type 2 quasar, 
but a large scale central dust lane (e.g., Thompson 1984) 
rather than a dusty torus could be responsible for the high dust 
extinction.  
Although the 8--13 $\mu$m spectrum of Cygnus A has been
presented by Ward (1996), neither the presence of the 9.7 $\mu$m
silicate dust absorption feature nor its optical depth are clear due to
limited signal-to-noise ratios.  
We conducted much more sensitive 8--13 $\mu$m spectroscopy to estimate 
the optical depth of the 9.7 $\mu$m silicate
dust absorption feature, thereby to investigate the location of
the obscuring dust.

\section{Observation and Data Analysis}

The 8--13 $\mu$m spectroscopy was conducted on the night of 1999, August 
21 (UT) at the Keck I Telescope using the Long Wavelength Spectrometer 
(LWS; Jones \& Puetter 1993)
under photometric sky conditions.  The seeing measured from a star was
$\sim0\farcs5$ in FWHM.  The LWS used a 128 $\times$ 128 Si:As array.
A low-resolution grating was used with a $0\farcs5$ wide slit and with
an N-wide filter (8.1--13 $\mu$m).  The resulting spectral
resolution was $\sim$50.

We utilized a ``chop and nod'' technique (e.g., Miyata et al. 1999) to
cancel the first order gradient of the sky emission variation and the
difference in background signal between different chopping beams.  The
frame rate was 20 Hz.  
Following the measurements at Mauna Kea by Miyata et al. (1999), 
the chopping and nodding frequencies were set to 5 Hz and 1/30 Hz, 
respectively, to achieve a background-limited sensitivity. 
In the actual data, a background-limited sensitivity may
not have been achieved, since stripe-like noise patterns were
recognizable in the array.  Since the emission region at 10 $\mu$m 
of Cygnus A and the standard star Vega (=$\alpha$ Lyrae, =HR7001) were
spatially unresolved, $\sim0\farcs5$ in FWHM, we set
the chopping amplitude at 3$''$ so as to maximize the observing
efficiency by placing the objects on the array all the time in both
the chopping and the nodding beams.  The total on-source integration
time of Cygnus A was 1650 sec.  Vega was observed just before the
Cygnus A observation, with an air mass difference less than 0.1, and
was used as a spectroscopic standard star.

We followed a standard data analyzing procedure, using IRAF 
%-------------
\footnote{
IRAF is distributed by the National Optical Astronomy Observatories, 
which are operated by the Association of Universities for Research 
in Astronomy, Inc. (AURA), under cooperative agreement with the 
National Science Foundation.}.  
%-------------
We first defined high dark current pixels and low-sensitivity pixels 
using, respectively, the dark and blackbody frames taken just after 
the Cygnus A observation.  
We replaced the data of these pixels with the interpolated signals 
of the surrounding pixels.  
The slit positions were nearly the same for Cygnus A and Vega, 
but the signal positions on the slit were slightly different.  
We corrected for the pixel variation of quantum efficiency along 
the slit by using the blackbody frame, whose flux was uniform along 
the slit. 
We extracted the spectra of the target and the standard star 
using an optimal extraction algorithm.

The wavelengths were calibrated using sky lines.  The Earth's
atmospheric transmission shows small and narrow dips at 11.7 $\mu$m
and 12.6 $\mu$m (Tokunaga 1998) that are easily discernible in raw
images as local maxima of sky background emission.  We examined the
positions of these two local maxima and confirmed that the wavelength
per pixel was 0.043 arcsec pixel$^{-1}$, consistent with the designed
value.  We therefore calibrated the wavelengths by assuming a linear
relationship between wavelength and pixel of 0.043 arcsec pixel$^{-1}$
throughout the detector.  In the wavelength-calibrated data, a
broad local maximum of sky background emission is found at 9.3--9.7
$\mu$m.  This wavelength range corresponds to the broad local minimum
of the Earth's atmospheric transmission (Tokunaga 1998).  This
wavelength calibration is believed to be accurate to within 0.05
$\mu$m.

Although Vega is known to show an infrared excess at $>$20 $\mu$m, 
the excess is not appreciable at $<$15 $\mu$m
(Heinrichsen, Walker, \& Klaas 1998).
We divided the signals of Cygnus A by those of Vega and
multiplied the result by a blackbody profile of 9400 K 
(Cohen et al. 1992).  
In our spectroscopy, because the slit width was comparable to 
the seeing, some signal was
possibly lost due to slight tracking errors; thus a high ambiguity in
flux calibration may be introduced.  We calibrated the flux of
Cygnus A in such a way that our spectrum between 8.1 $\mu$m and 13
$\mu$m agreed with the $N$-band photometric data ($N$ = 0.18 Jy or 5.7
mag; Rieke \& Low 1972; Heckman et al. 1983).
The flux measurement using our data agrees with this 
by a factor of $\sim$2.

\section{Results}

The flux-calibrated spectrum of Cygnus A is shown in Figure 1.  A
broad absorption-like feature is seen at a peak wavelength of $\sim$10
$\mu$m.  This wavelength is consistent to the peak wavelength of the
9.7 $\mu$m silicate dust absorption feature redshifted with {\it z} =
0.056.  Hence, the broad absorption-like feature is likely to
originate from silicate dust absorption.

It has been suggested that, if a galaxy is powered strongly by
star-formation activities, and the polycyclic aromatic hydrocarbon
(PAH) emission features at 7.7 $\mu$m, 8.6 $\mu$m, and 11.3 $\mu$m are
strong, then the mid-infrared spectra mimic those with 9.7
$\mu$m silicate dust absorption (Genzel et al. 1998).  In the case of
Cygnus A, however, the extinction-corrected 2--10 keV hard X-ray
luminosity ($L_{\rm X}$(2--10 keV) $=$ 1--5 $\times$ 10$^{44}$ ergs
s$^{-1}$) relative to the 40--500 $\mu$m far-infrared luminosity
($L_{\rm FIR}$ $=$ 0.7--1.6 $\times$ 10$^{45}$ ergs s$^{-1}$)
%--------------
\footnote{ We use the formula $L_{\rm FIR}$ = 2.1 $\times$ 10$^{39}$
$\times$ D(Mpc)$^{2}$ $\times$ (2.58 $\times$ $f_{60}$ + $f_{100}$),
where $f_{60}$ and $f_{100}$ are, respectively, the {\it IRAS} 60
$\mu$m and 100 $\mu$m flux in Jy (Sanders \& Mirabel 1996).  The
$f_{60}$ and $f_{100}$ fluxes of Cygnus A are 2.329 Jy and $<$8.278 Jy, 
respectively.  }
%--------------
is ${\ ^{\displaystyle >}_{\displaystyle \sim}\ }$0.1, 
as high as in galaxies powered predominantly by AGN activity.  
Furthermore, the wavelength coverage of our spectrum
(8.1--13.0 $\mu$m) is 7.7--12.3 $\mu$m in the rest-frame and hence
covers the 11.3 $\mu$m PAH emission feature (10.9--11.6 $\mu$m in the
rest-frame; Rigopoulou et al. 1999).  We find no detectable 11.3
$\mu$m PAH emission feature at 11.9 $\mu$m in the observed frame ($<$6
$\times$ 10$^{-17}$ W m$^{-2}$ in flux or $<$4 $\times$ 10$^{41}$ ergs
s$^{-1}$ in luminosity).  The 11.3 $\mu$m PAH to far-infrared 40--500
$\mu$m luminosity ratio is $<$6$\times$10$^{-4}$, more than an order
of magnitude smaller than that found for galaxies powered by
star-formation activities (0.009$\pm$0.003; Smith, Aitken, \& Roche
1989).  Hence, it is very unlikely that the shorter side of our
mid-infrared spectrum is dominated by PAH emission features from
star-formation activities.  We therefore ascribe the broad
absorption-like feature fully to 9.7 $\mu$m silicate dust absorption.

\section{Discussion} 

\subsection{The Small Optical Depth of Silicate Dust Absorption}

Since the Galactic dust extinction toward the Cygnus A nucleus is 
estimated to be small, $A_{\rm V}$ $\sim$ 1 mag 
(Spinrad \& Stauffer 1982; van den Bergh 1976), 
the 9.7 $\mu$m silicate dust absorption is believed to be attributed 
to dust in the Cygnus A galaxy. 
If we adopt the ratio of visual extinction to the 
optical depth of the 9.7 $\mu$m silicate dust absorption 
found in the Galactic interstellar medium 
($A_{\rm V}$/$\tau_{9.7}$ $=$ 9--19; Roche \& Aitken 1985),
a dust extinction of $A_{\rm V}$ $\sim$ 150 mag should 
provide $\tau_{9.7}$ = 7.9--16.6.  
In this case, the flux at the absorption peak should be attenuated 
by a factor of 2.7$\times$10$^{3}$ -- 1.6$\times$10$^{7}$, 
and hence should be saturated in the spectrum.
The observed spectrum, however, shows no such saturation at all.
If we use the following formula by Aitken \& Jones (1973), 
\begin{center}
$\tau_{9.7}$ $=$ ln [$\frac{F_{\rm \lambda}(8) + F_{\rm \lambda}(13)}
{2 \times F_{\rm \lambda}(9.7)}$] (rest-frame), 
\end{center} 
then $\tau_{9.7}$ is only $\sim$1, much smaller than that expected 
from $A_{\rm V}$ $\sim$ 150 mag.  
This is the predicted trend when obscuring dust exists so close 
to a central engine that a temperature gradient occurs.  
Before pursuing this possibility in more detail, we
review another possibility that could explain the small $\tau_{9.7}$.

Interstellar dust consists mainly of silicate and carbonaceous dust
(Mathis, Rumpl, \& Nordsieck 1977; Mathis \& Whiffen 1989).  If the
contribution of silicate dust to $A_{\rm V}$ is much smaller than that
in the Galactic interstellar medium, $\tau_{9.7}$ could be small even
in the case of high $A_{\rm V}$.  
However, since carbonaceous dust is more fragile than silicate dust 
(Draine \& Salpeter 1979), carbonaceous dust should have been more 
destroyed than silicate dust around the Cygnus A nucleus, where 
a radiation field is expected to be much stronger and more energetic 
than that in the Galactic interstellar medium. 
The depletion of silicate dust relative to carbonaceous dust 
around Cygnus A is very unlikely.  
We therefore argue that a temperature
gradient is the most likely explanation that could reproduce 
the observed small $\tau_{9.7}$.

\subsection{Modeling}

In this subsection, we examine whether a temperature gradient can
{\it quantitatively} explain the small $\tau_{9.7}$ in spite of a high
$A_{\rm V}$ toward the background $L$-band emission region.  In
addition to our mid-infrared spectrum, we incorporate data points at
3--30 $\mu$m to constrain our model parameters. 
This is because dust emission
powered by obscured AGN activity is strong at 3--30 $\mu$m and thus 
emission in this wavelength range can provide useful information on 
the dust distribution around a central engine.  
The photometric data at 3--30 $\mu$m used are summarized in Table 1.  
We do not incorporate data at $<$3 $\mu$m and at $>$30 $\mu$m, 
because stellar emission dominates the flux of Cygnus A 
at $<$3 $\mu$m (Djorgovski et al. 1991), while emission 
from cold dust in the host galaxy could contribute 
significantly at $>$30 $\mu$m.

We estimate the contribution of (1) synchrotron emission and (2)
stellar emission to the 3--30 $\mu$m emission of Cygnus A.  
Firstly, the spectral energy distribution of the Cygnus A nucleus 
(without hotspots emission) shows a clear flux excess in the 
infrared regions compared to the extrapolation of the synchrotron 
emission component at longer wavelengths (Haas et al. 1998).
In fact, if we extrapolate the data at 0.45--2.0 mm 
using F$_{\nu}$ $\prec$ $\nu^{-0.6}$ (Robson et al. 1998) and 
assume an extinction of $A_{\rm V}$ $=$ 150 mag toward
the synchrotron emission region of the Cygnus A nucleus, then 
we find the synchrotron emission component contributes less 
than 1/10 of the observed flux at 3--30 $\mu$m.  
Secondly, the contribution from the stellar emission can be
estimated from the near-infrared $K$-band (2.2 $\mu$m) photometric data
($K$ = 13.78$\pm$0.06 mag), which are dominated by stellar emission in
the case of Cygnus A (Djorgovski et al. 1991).  
If we assume the $K-L$ and $K-L'$ colors of late type stellar 
populations in normal galaxies (0.2--0.4; Willner et al. 1984), 
the stellar contribution at $L$ and $L'$ is 13.4--13.6 mag.  
This is smaller than $\sim$30\% of the observed flux at $L$ and $L'$.
At wavelengths longer than $L'$, the stellar contribution decreases 
and becomes negligible.

As a consequence, the 3--30 $\mu$m data of Cygnus A should be
dominated by dust emission powered by obscured AGN activity.  We use
the code {\it DUSTY} developed by Ivezic, Nenkova, \& Elitzur (1999)
to investigate whether the observed small $\tau_{9.7}$ as well as the
spectral energy distribution at 3--30 $\mu$m can be quantitatively
explained by emission from dust in the vicinity of a central AGN.

The code {\it DUSTY} solves the radiative transfer equation for a
source embedded in a spherically symmetric dusty envelope. It creates
an output spectrum as the sum of the attenuated input radiation, dust
emission, and scattered radiation.  For Cygnus A, the presence of
radio hot spots (Wright \& Birkinshaw 1984), the detection of strong
optical [OIII] emission (Osterbrock \& Miller 1975), and
centrosymmetric polarization patterns (Tadhunter, Scarrott, \& Rolph
1990; Ogle et al. 1997) indicate that the dusty envelope is
torus-like, with dust along an unknown direction roughly perpendicular
to our line of sight expelled with an unknown solid angle.  Hence,
strictly speaking, the assumption of spherical symmetry is not valid.
If we view the torus-like structure from a face-on direction where the
unattenuated hot ($\sim$1000 K) dust emission from the innermost
envelope is seen directly, the output dust emission spectrum is
expected to be quite different to the one from a spherical dusty
envelope around a central engine, because all the hot dust emission is
attenuated in the spherical model.  However, it is most likely in the case
of Cygnus A that we are viewing the torus
from almost an edge-on direction, where
no unattenuated emission from the innermost envelope is seen.
Furthermore, because our aim is to explain the small $\tau_{9.7}$, the
presence of a temperature gradient in the obscuring dust along our
line of sight in front of a background emitting source is the most
important factor.  The geometry of dust perpendicular to our line of
sight would not be crucial.  Hence, for simplicity, we assume a
spherically symmetric dust envelope.

Thanks to the general scaling properties of the radiative transfer
mechanism (Rowan-Robinson 1980; Ivezic \& Elitzur 1997), there are
only several free parameters.  Firstly, we assume the input radiation
as blackbody with a temperature of 40000 K, following Rowan-Robinson
\& Efstathiou (1993).  This is not very critical as explained by 
Rowan-Robinson (1980).  Next, we set the dust sublimation temperature
as 1000 K following Rowan-Robinson \& Efstathiou (1993) and Dudley \&
Wynn-Williams (1997).  Since, in this model, the extinction estimated
using $\sim$3 $\mu$m data is believed to be nearly the same as 
the extinction toward the central engine ($\S$ 1), 
we set the optical depth at 0.55
$\mu$m ($\sim$ $A_{\rm V}$/1.08) toward the central engine to $\tau$
$=$ 140.  We adopt the standard dust size distribution by Mathis et
al. (1977) and standard interstellar dust mixture as defined in {\it
DUSTY}.  The ratio of the outer to the inner radius of the dusty
envelope is set as a free parameter.  We only try a simple power law
radial density profile ($\prec$ r$^{\gamma}$), and the value $\gamma$
is also set as a free parameter.

When searching for parameters that can fit the observed data, we allow
the model output spectrum and the observed data to differ by a few
factors, particularly at shorter wavelengths.  This is because the
period over which the data in Table 1 were taken spans 10 years.
Since emission at shorter wavelengths is from warm dust located at the
inner part of a dusty envelope, variability over a time scale of 10
years could be significant.

In Figure 2, we show the results of a calculation that provides a
reasonable fit to the observed data, particularly at longer wavelengths.  
The outer-to-inner radius ratio is 200, and the power-law index 
of dust radial distribution is 2.5.
If we adopt a value of $\sim$10$^{45}$ ergs s$^{-1}$ as the
luminosity of the central radiation source, the physical scales of
the inner radius of the dusty envelope is 2.25 pc.
An outer-to-inner radius ratio of 80--500 and a power-law index 
of 2.5--3.0 for the dust radial distribution produce 
an output spectrum similar to that shown in Figure 2.

In summary, the observed small $\tau_{9.7}$ as well as the spectral
energy distribution at 3--30 $\mu$m of Cygnus A can be quantitatively
explained by a model in which the emission at 3--30 $\mu$m originates
from thermal emission by dust in the vicinity of a central AGN.
Combining this result with the huge AGN luminosity of Cygnus A
($L_{\rm X}$(2--10 keV) $>$ 10$^{44}$ ergs s$^{-1}$), we argue that
observed data are consistent with the picture of Cygnus A being a
type 2 quasar, that is, a highly luminous AGN that is highly obscured
($A_{\rm V}$ $>$ 50 mag) not by $>$ a few 100 pc scale dust in 
the host galaxy, but by a dusty torus with an inner radius of $<$10 pc.

\subsection{Another type 2 quasar}

Although we demonstrated that the observed data are
consistent with the picture of Cygnus A being a type 2 quasar, 
Cygnus A is a radio-loud AGN (the minor AGN population) and 
is a cD galaxy at the center of a cluster of galaxies 
(Spinrad \& Stauffer 1982).  
It may be thought that Cygnus A is an unusual example.
In this subsection, we mention briefly the emission properties of 
a radio-{\it quiet} AGN (the major AGN population), IRAS 08572+3915.

IRAS 08572+3915 ({\it z} = 0.058) is one of the nearby ultra-luminous
infrared galaxies ($L_{\rm IR}$ $>$ 10$^{12}$ $L_{\odot}$; Kim,
Veilleux, \& Sanders 1998).  The radio 20 cm to the far-infrared
40--500 $\mu$m flux ratio is similar to those of radio-quiet AGNs
(Crawford et al. 1996).  It displays a strong absorption-like feature
at 10 $\mu$m (Dudley \& Wynn-Williams 1997).  
Although the interpretation of the absorption-like feature at 
$\sim$10 $\mu$m is sometimes difficult ($\S$ 3), 
this source displays a very strong 3.4 $\mu$m carbonaceous dust 
absorption feature and no detectable 3.3 $\mu$m PAH emission 
feature in the 3--4 $\mu$m spectrum (Wright et al. 1996), 
strongly suggesting that IRAS 08572+3915 is powered by a highly 
embedded AGN and not by star-formation activities
%--------------
\footnote
{Dudley \& Wynn-Williams (1997) argue that the 8--22
$\mu$m spectrum of Arp 220 and that of IRAS 08572+3915  
share similar properties.  However, another interpretation of 
the 10 $\mu$m spectrum of Arp 220 has been proposed by 
Genzel et al. (1998).  Given the absence of a
high-quality 3--4 $\mu$m spectrum of Arp 220, it is unknown which
interpretation is correct.  In this paper we tentatively regard only
IRAS 08572+3915 as a convincing sample of a type 2 quasar.}. 
%--------------
Hence, the strong absorption-like feature at $\sim$10 $\mu$m must be
fully attributed to 9.7 $\mu$m silicate dust absorption.  Based on the
optical depth ratio between the 9.7 $\mu$m and 18 $\mu$m silicate dust
absorption features, the presence of a temperature gradient in the
obscuring dust around a very compact energy source (= AGN) has been
argued (Dudley \& Wynn-Williams 1997).  
If we adopt $\tau_{9.7}$ $=$ 5.2 (Dudley \& Wynn-Williams 1997) 
and the Galactic optical depth ratio of the 3.4 $\mu$m carbonaceous 
to the 9.7 $\mu$m silicate dust absorption 
($\tau_{3.4}$/$\tau_{9.7}$ $=$ 0.06--0.07; Pendleton et al. 1994; 
Roche \& Aitken 1985), then $\tau_{3.4}$ $\sim$ 0.35 is expected.  
The observed $\tau_{3.4}$ is $\sim$ 0.9 (Pendleton 1996), 
more than a factor of 2 higher than expected.  
This means the extinction estimate at $\sim$3 $\mu$m is higher than
that at $\sim$10 $\mu$m and thus supports the presence of a temperature 
gradient in the obscuring dust around the AGN.
All the results are consistent with the picture of IRAS 08572+3915 
being a radio-quiet type 2 quasar.

IRAS 08572+3915 has a LINER-type optical spectrum (Kim et al. 1998).
No broad ($>$ 2000 km s$^{-1}$ in FWHM) emission line has been
detected in the 2 $\mu$m spectrum (Veilleux, Sanders, \& Kim 1999).
The detection of 2--10 keV hard X-ray flux by the Advanced Satellite 
for Cosmology and Astrophysics ({\it ASCA}) is at most marginal 
(based on our quick look into the {\it ASCA} archive).
Namely, no sign of strong AGN activity has been detected at $<$2
$\mu$m or even at hard X-ray.  The weak hard X-ray flux is not
surprising, given the high obscuration toward the nucleus.  The
estimated $A_{\rm V}$ toward the 3 $\mu$m emission region (that is,
toward the innermost part of the obscuring dust) from the observed
$\tau_{3.4}$ is 130--220 mag, if we adopt the relation
$\tau_{3.4}$/$A_{\rm V}$ $=$ 0.004--0.007 found in the Galactic
interstellar medium (Pendleton et al. 1994).  If the dust-to-gas ratio
is more than a factor of five higher than the Galactic value 
($N_{\rm H}$/$A_{\rm V}$ = 1.8 $\times$ 10$^{21}$ cm$^{-2}$ mag$^{-1}$), 
as observed in some AGNs (see $\S$ 1), 
then $N_{\rm H}$ is higher than 10$^{24}$ cm$^{-2}$, 
and direct 2--10 keV hard X-ray emission is completely blocked.

Unlike Cygnus A, the signs of a type 2 quasar in IRAS 08572+3915 
were first recognized through study of thermal infrared regions. 
We suggest that other type 2 quasars that are not recognizable at $<$2
$\mu$m and at hard X-ray region may also be found through  
detailed study of the thermal infrared region.

\section{Summary}  

Our main results are the following.

\begin{enumerate}
\item We detected a 9.7 $\mu$m silicate dust absorption feature
toward the Cygnus A nucleus.
\item The optical depth of the absorption feature ($\tau_{9.7}$ 
      $\sim$ 1) is smaller by a large factor than that expected 
      from $A_{\rm V}$ toward the background $L$-band emission region 
      ($\sim$150 mag).
\item We demonstrated that the small optical depth together with the
      spectral energy distribution at 3--30 $\mu$m could be 
      quantitatively explained by emission from a dusty envelope 
      in the vicinity ($<$10 pc in inner radius) of a central AGN.
\item Combining this finding with the huge AGN luminosity of
      Cygnus A, we argued that the observed data are consistent with 
      the picture of Cygnus A being a type 2 quasar.
\end{enumerate}

\acknowledgments      
         
We thank R. Campbell and T. Stickel for their support during 
the Keck observing run, Dr. C. C. Dudley for his useful comments 
on the manuscript, and L. Good for her proofreading of this paper.
Drs. A. T. Tokunaga and H. Ando support my stay at the University of 
Hawaii.
MI is financially supported by the Japan Society for the Promotion 
of Science during his stay at the University of Hawaii.

\clearpage

\newpage

\begin{center}
\begin{deluxetable}{ccc}
\tablewidth{6.5in}
\tablenum{1}
\tablecaption{Photometric Data of Cygnus A at 3--30 $\mu$m}
\tablecolumns{3}
\tablehead{
\colhead{Wavelength}& \colhead{Flux}& \colhead{Reference} \\
\colhead{($\mu$m)}&  \colhead{(Jy)} & \colhead{} 
}
\startdata
3.55 ($L$) & (8.5$\pm$1.0)$\times$10$^{-3}$ (11.28$^{+0.14}_{-0.12}$ mag)
& Heckman et al. (1983)\\
3.75 ($L'$) & (3.8$^{+1.6}_{-1.2}$)$\times$10$^{-3}$ 
(12.05$\pm$0.4 mag)  & Djorgovski et al. (1991) \\
4.8  & $<$0.030 & Haas et al. (1998) \\
7.3  & 0.061    & Haas et al. (1998) \\
12   & $<$0.250 & $IRAS$ Point Source Catalog \\
12.8 & 0.485    & Haas et al. (1998) \\
20   & 0.816    & Haas et al. (1998) \\
25   & 1.062    & $IRAS$ Point Source Catalog  \\
\enddata

\end{deluxetable}
\end{center}

\clearpage

\begin{figure}[h]
\centerline{\psfig{file=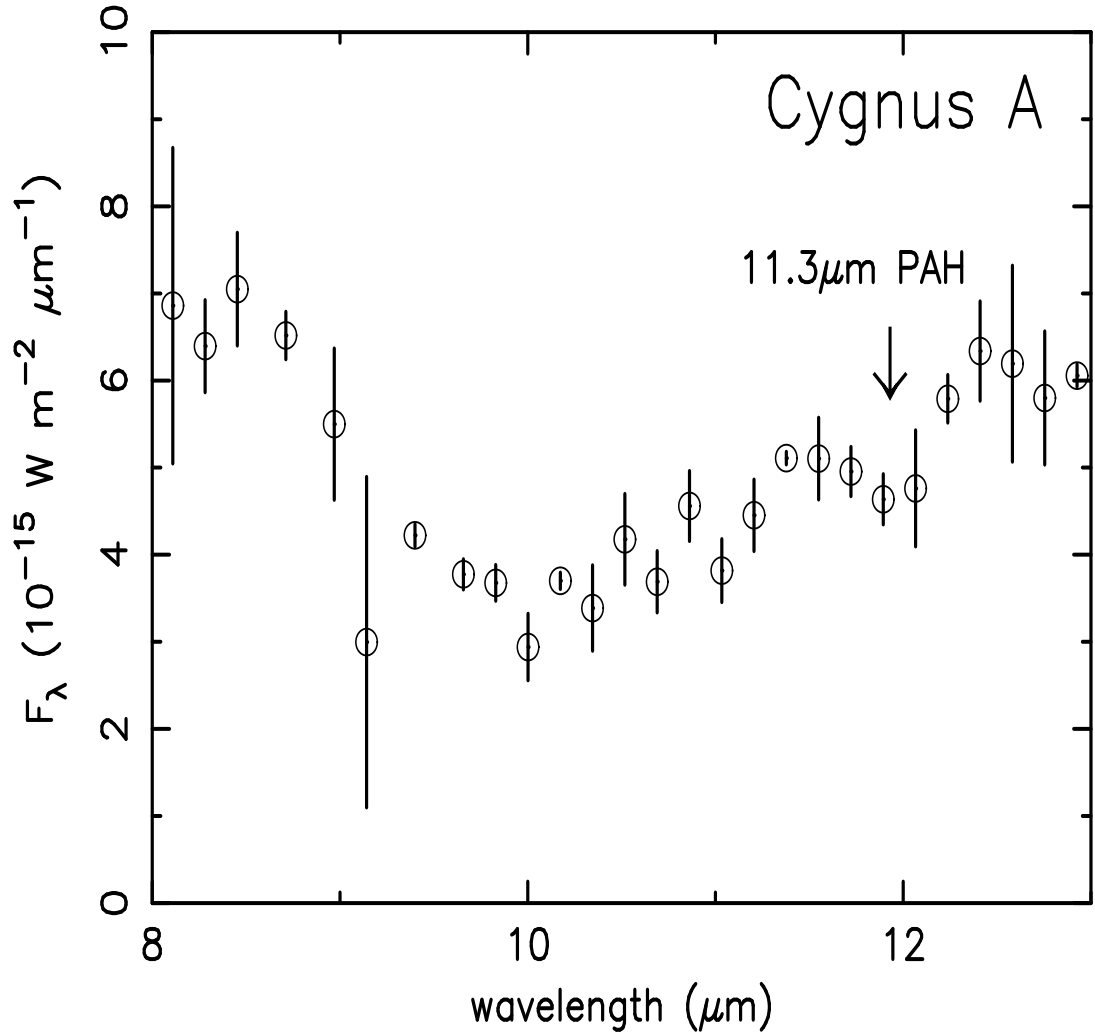,angle=0,width=6.5in}}
\caption{Flux-calibrated spectrum of Cygnus A.
The ordinate is $F_{\lambda}$ in W m$^{-2}$ $\mu$m$^{-1}$, and the 
abscissa is wavelength in $\mu$m in the observed frame. 
The down arrow indicates the wavelength of the redshifted 
11.3 $\mu$m PAH emission feature.}
\end{figure} 

\newpage

\begin{figure}[h] 
\centerline{\psfig{file=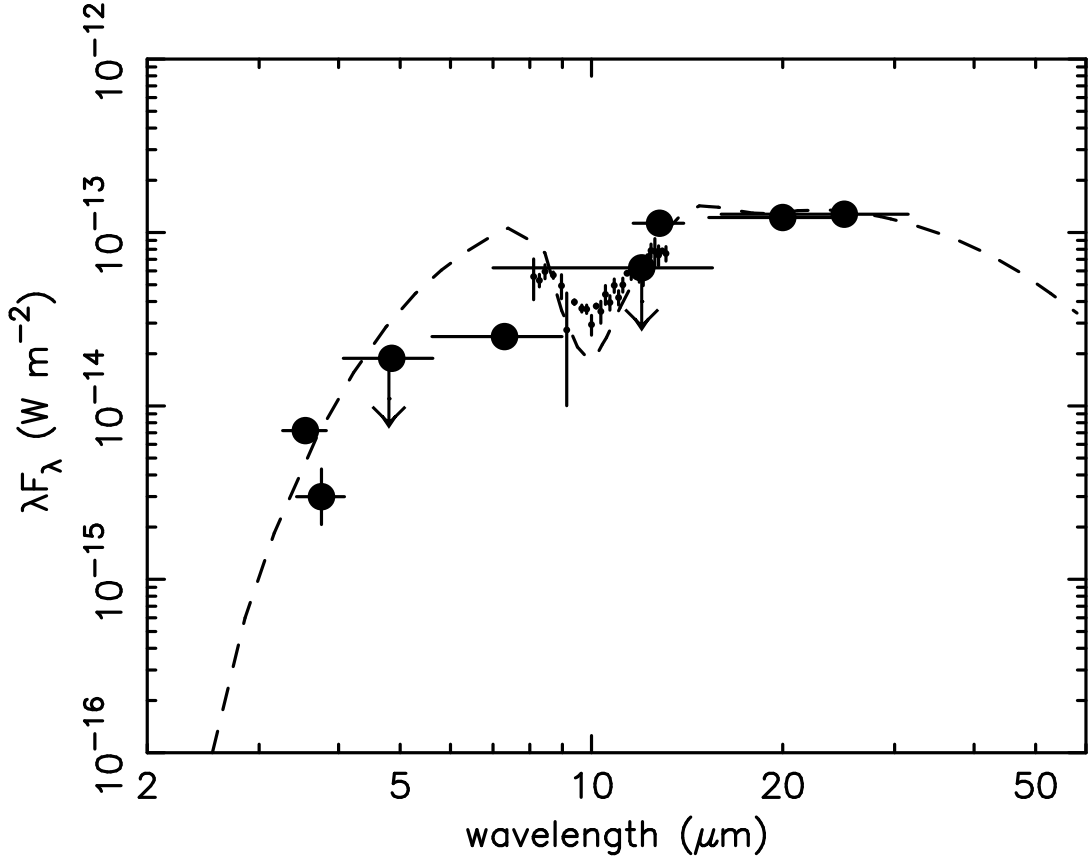,angle=0,width=6.5in}}
\caption{Spectral energy distribution of Cygnus A
at 3--30 $\mu$m.  The ordinate is $\lambda F_{\lambda}$ in W
m$^{-2}$, and the abscissa is wavelength in $\mu$m in the observed
frame.  The dashed line is the output spectrum of {\it DUSTY} with an
outer-to-inner radius ratio of 200, a power-law index of 2.5 in the
dust radial distribution, and an optical depth at 0.55 $\mu$m of 140
toward the central engine.  Large filled circles: Data from the
literature (Table 1).  Small filled circles: Our LWS data.
Any model output spectra that fit the observed data at $>$10 $\mu$m 
overpredict the fluxes at 4.8 and 7.3 $\mu$m. 
This could be caused by time variation (see $\S$ 4.2). 
}
\end{figure} 

\end{document}